\documentclass[prd,twocolumn, nofootinbib,noshowkeys,superscriptaddress]{revtex4}

\usepackage[latin1]{inputenc}
\usepackage{pstricks,pst-node,pst-text,pst-3d,pslatex}
\usepackage[T1]{fontenc}
\usepackage{amsmath}

\usepackage{graphicx,amsfonts}
\usepackage{amsmath}

\newcommand{\be}{\begin{eqnarray}}
\newcommand{\ee}{\end{eqnarray}}

\begin{document}
\title{Microscopically Computing Free-energy Profiles and Transition Path Time of Rare Macromolecular Transitions} 
\author{P. Faccioli}
\affiliation{Physics Department,  University of Trento, Via Sommarive 14 (Povo) I-38129, Trento (Italy).} 
\affiliation{INFN, Gruppo Collegato di Trento, Via Sommarive 14 (Povo) I-38129, Trento (Italy). }
\author{F. Pederiva}
\affiliation{Physics Department, University of Trento, Via Sommarive 14 (Povo) I-38129, Trento (Italy).} 
\affiliation{INFN, Gruppo Collegato di Trento, Via Sommarive 14 (Povo) I-38129, Trento (Italy). }
\begin{abstract}
We introduce a rigorous method to microscopically  compute the observables which characterize the thermodynamics and kinetics of rare macromolecular transitions for which it is possible to identify \emph{a priori} a slow reaction coordinate. 
In order to sample the ensemble of statistically significant reaction pathways, we define a biased molecular dynamics (MD)  in  which barrier-crossing transitions are accelerated  without introducing any unphysical external force. In contrast to other biased MD methods, in the present approach the systematic errors which are generated in order to accelerate the 
transition  can be analytically calculated and therefore can be corrected for.
This  allows for a computationally efficient reconstruction of the free-energy profile as a function of the reaction coordinate and for the calculation of the corresponding diffusion coefficient. 
The transition path time can then be readily evaluated within the Dominant Reaction Pathways (DRP) approach. We illustrate and test this method 
by characterizing a thermally activated transition on a two-dimensional energy surface and the folding of a small protein fragment within a coarse-grained model.   
\end{abstract}
\maketitle

\section{Introduction}

The recent developments in single-molecule optical- and force- spectroscopy allow to experimentally characterize the thermodynamics and kinetics of many fundamental biomolecular reactions to an unprecedented level of accuracy. For example, pulling experiments based on optical tweezers~\cite{tweezers} or atomic-force microscopy \cite{AFM} can
 provide the full free-energy profile of biopolymers as a function of their end-to-end distance~\cite{reconstr}, 
while the single-molecule F\"oster Resonance Energy Transfer spectroscopy yields the reaction rate~\cite{FRET} and, very recently, the transition path time (TPT)~\cite{TPT}.  

  The possibility of measuring these observables poses the challenge to predict their dynamics from microscopic atomistic simulations. 
 Unfortunately, the MD algorithms are very inefficient to this purpose, because they require to simulate time intervals which are  exponentially long 
in the free-energy barrier. 
 
These limitations have motivated the development of alternative theoretical frameworks to investigate the free energy landscape \cite{eq1,eq2,eq3} and reaction 
kinetics properties \cite{neq1,neq2,neq3,neq4,neq5,neq6,neq7, neq8,neq9,neq10} of activated reactions. Some of these methods --- such as e.g. the
meta-dynamics approach \cite{eq1}--- involve a suitable choice of a set of reaction coordinates which are used to bias and accelerate 
the exploration of the energy landscape.  By contrast, methods like transition interface sampling \cite{neq8}, milestoning \cite{neq9} or dynamics Monte 
Carlo  \cite{neq10} sample directly the space of reactive pathways, without introducing a bias on the dynamics. 
On the other hand, these methods are in general computationally quite costly.

 In a recent work,  a variant of the Dominant Reaction Pathway (DRP) method \cite{DRP0, DRP1,DRP2} has been 
  developed~\cite{DRPpnas}, which generates statistically significant protein folding pathways by combining an accelerated  MD algorithm  \cite{ratchetMD}  with a path-integral based variational approach. 
Using the accelerated MD, several hundreds of folding trajectories for single-domain proteins of typical size can be generated 
in just a few hundreds of CPU hours. 
The trial paths are then ranked in terms of their statistical weight in the  (unbiased) over-damped Langevin dynamics, and the most probable (i.e. least
 biased) trajectories among them are identified. This way, many of such so-called dominant reaction pathways, each corresponding to a different initial condition, have been computed 
   for a WW  protein domain using a realistic force field~\cite{DRPpnas}. These paths were found to agree very well with those obtained by Shaw and co-workers within the same force field, by means of ultra-long MD simulations on the Anton special-purpose machine~\cite{Anton}. 

The high  efficiency of the DRP method comes from the fact that its computational time does not scale exponentially with the free-energy barriers. By this method it is now possible to atomistically study the dynamics of polypeptides with realistic size and kinetics, and even simulate the folding of complex knotted proteins~\cite{knotted}.  

The main limitation of this approach is that, since the accelerated dynamics used to generate the ensemble of reaction pathways  breaks microscopic reversibility, it cannot be used to directly compute kinetic and thermodynamic observables. 
Due to this problem, to date the DRP method has been used only to investigate the reaction mechanism, for example by characterizing the transition state ensemble.

In this work, we overcome this limitation. We devise a rigorous scheme to  compute the 
transition path time and evaluate the potential of mean-force of a previously determined slowly-evolving collective coordinate (CC). The method is based on a new type of accelerated MD algorithm,
herby called \emph{hindered molecular dynamics} (hMD). In contrast to other methods, in the hMD algorithm, no external force is introduced to speed up the reaction. 
In addition, the effect of the bias on the evolution of a slow reaction coordinate can be rigorously and analytically computed, hence can be corrected for. As a result, it is possible to extract the potential of mean-force and the diffusion coefficient of the reaction coordinate from a set of suitable averages evaluated over the reactive trajectories obtained from the biased dynamics.  Once these quantities have been determined, the transition path time can be easily computed employing the DRP formalism. 

The paper is organized as follows. In the next section  we introduce the hMD dynamics and show how to extract the potential of mean-force and the diffusion constant of a slow reaction coordinate. In section \ref{TPT} we discuss how to compute the transition path time in the DRP formalism. Section \ref{tests} provides two illustrative applications to systems of increasing complexity. Finally, results and  conclusions are summarized in section \ref{conclusions}.

\section{Computing the potential of mean-force and the diffusion coefficient of a CC from hMD simulations}

Let us begin by considering the over-damped Langevin equation which mimics  the microscopic dynamics of the molecule in a solvent.
 In the so-called Ito calculus this equation is defined as:
\be
\label{langevin}
x_{i+1} = x_i -(\Delta t/\gamma) \nabla U(x_i) + \sqrt{2D\Delta t} ~\eta_i, 
\ee
where $\gamma$ is the viscosity,  $D=\frac{k_BT}{\gamma}$ is the diffusion coefficient, $x_i$ is the point in the $3N$-dimensional configuration space visited at the $i-$th time step,  $U(x)$ is the potential energy and $\eta(t)$
is a stochastic variable sampled from a Gaussian distribution with 
with zero average and unitary variance. The high-friction limit which underlies the over-damped Langevin Eq. (\ref{langevin}) is appropriate for many systems of biophysical interest. For example, in proteins, the effects of the acceleration affects the dynamics only at time scales smaller than few fractions of ps \cite{pitard}.

We now introduce the hMD,  which is closely related to the ratchet-and-pawl MD\cite{ratchetMD} used
to generate trial paths in our previous protein folding DRP calculations~\cite{DRPpnas}.
The main difference between the two algorithms is that, in the hMD, no unphysical external force is introduced to disfavor fluctuations in the direction of the reactant. Instead,  when the system tries to evolve backwards along a reaction coordinate,   the dynamics is slowed down by increasing the viscosity and the decreasing the heat-bath temperature. 

Namely, denoting with  $z(x)$ a  configuration-dependent CC ---assumed for definiteness to 
monotonically increase from the reactant to the product---, 
the hMD is defined  by the following stochastic differential equation:
\be
\label{hMD}
x_{i+1} &=& \theta\left[ z(x_{i+1})-z(x_{i})\right]~\left( x_i -\frac{\Delta t}{\gamma} \nabla U(x_i) + \sqrt{2 D \Delta t} \eta_i\right) \nonumber\\
 &+&\theta\left[z(x_{i})- z(x_{i+1})\right]~\left( x_i -\frac{\Delta t}{\xi \gamma} \nabla U(x_i) + \frac{1}{\xi} \sqrt{2 D \Delta t}  \eta_i\right),\nonumber\\
\ee
where $\xi>1$ is called the hindering coefficient.

By scoring the trajectories generated by the hMD (\ref{hMD}) according to the path probability of the unbiased Langevin dynamics (\ref{langevin}) one can efficiently 
obtain an ensemble of statistically representative reaction pathways, see Ref. \cite{DRPpnas}. Unfortunately, the time intervals of the hMD (\ref{hMD}) are not physically meaningful and dominant pathways alone do not allow to compute free-energy differences. 

In order to overcome this problem and establish the connection with kinetics and thermodynamics our strategy is to analyze the average time evolution of some slow reaction coordinate $Q$.  In general, such a collective variable does not necessarily need to coincide with the biasing variable~$z$, but can be related to it by a constant scaling factor, which produces a rescaling of the diffusion coefficient. 

In the limit in which the spontaneous time evolution of the CC $Q$ very slow compared to that of all microscopic degrees of freedom, its  (unbiased) dynamics  can be described by an effective over-damped Langevin equation:
\be\label{effLangevin}
Q_{i+1} = Q_i -(\Delta t/\gamma_Q)~ G'(Q_i) + \sqrt{2 k_BT \Delta t/\gamma_Q} ~\eta_i, 
\ee 
where $\gamma_Q$ and $D_Q=k_BT/\gamma_Q$ the respectively the viscosity and the diffusion coefficient of the CC. In the following we restrict to the case in which the diffusion constant is assumed to be state independent (white noise). The generalization to colored noise is possible but it requires a careful choice of the stochastic calculus, and will not be considered in this first work. 
 
In an hMD simulation,  any time the system evolves towards a smaller value of the $Q$  (hence of $z$), the dynamics is slowed down
 by the same rescaling of the diffusion coefficient and  temperature of the underlying microscopic dynamics.  Hence, the equation of motion of $Q$ in a hMD simulation is given by:
 \be
 \label{sde}
Q_{i+1}-Q_i &=&  \left( -\frac{ \Delta t }{\gamma_Q}~ G'(Q_i)+\sqrt{2 D_Q \Delta t}~\eta_i\right) \\
&\cdot&\left\{1+\left(\frac{1}{\xi}-1\right)\theta\left[\sqrt{\frac{\Delta t}{2 \gamma_Q k_B T}}~ G'(Q_i)- \eta_i \right]\right\},\nonumber
\ee
where the role of the step-function is to hinder the dynamics when the fluctuation would drive the reaction backwards. 

From the stochastic differential Eq. (\ref{sde}) it is immediate to compute
 the probability for  the system to  evolve
  from a configuration with CC $Q_i$ to one with CC $Q_{i+1}$, in a elementary step $\Delta t$ of hMD simulation:
\be\label{Pz}
&&\mathcal{P}(Q_{i+1}, \Delta t|Q_i) = \mathcal{N}~\left(
 \xi e^{\frac{-\xi^2(\Delta Q + \frac{\Delta t}{\xi \gamma_Q}G'(Q_i))^2}{4 D_Q\Delta t}}
\theta\left[-\Delta Q \right]\right.\nonumber\\
&& \left. +  e^{\frac{-(\Delta Q  + \frac{\Delta t}{\gamma_Q}G'(Q_i))^2}{4 D_Q \Delta t}}
~\theta\left[\Delta Q  \right]\right),
\ee
where  $\Delta Q \equiv Q_{i+1}-Q_i$ and $\mathcal{N}= \sqrt{\frac{\gamma_Q}{4 \pi ~\Delta t ~k_B T}}$.

Using this equation we  compute the average infinitesimal 
displacement of the CC  $Q$ in a time interval $\Delta t$ of hMD, starting from configurations
in which the CC takes a value $Q$: 
  \be\label{main1}
   \langle \Delta Q(Q) \rangle_{hMD} = \sqrt{\frac{\Delta t k_B T}{\pi \gamma_Q}}\frac{\xi-1}{\xi}
   - \frac{\Delta t}{2 \gamma_Q} G'(Q) \frac{1+\xi}{\xi} +\ldots
   \ee
   Similarly, the average square displacement  $\langle \Delta Q^2(Q)\rangle_{hMD}$ reads:
\be\label{main2}
\langle \Delta Q^2(Q)\rangle_{hMD} &=& \frac{\Delta t k_B T}{\gamma_Q}~\frac{1+\xi^2}{\xi^2}+\ldots 
\ee
In both equations (\ref{main1}) and (\ref{main2}) the dots denote  corrections of order  $\Delta t^{3/2}$.
If the biasing coordinates is defined in such a way to decrease (rather than increase) from the reactant to the product, the first term in the right-hand-side of 
Eq. (\ref{main1}) changes  sign. 
The averages in Eq.s (\ref{main1}) and (\ref{main2}) can be efficiently evaluated, 
hence allowing  to determine $\gamma_Q$ and $G(Q)$. 

\section{Computing the Transition Path Time}
\label{TPT}  
In the previous section we have discussed how it is possible to compute the potential of mean-force $G(Q)$ and the diffusion coefficient $D_Q$, by evaluating averages over microscopic trajectories obtained in the hMD.  We now show that, once these quantities have been determined, it is possible to restore the correct time scales 
in the calculated reactive trajectories, by applying the DRP formalism to the stochastic projected dynamics of the CC defined in Eq. (\ref{effLangevin}).

The starting point of the DRP approach is the path integral representation of the conditional probability of going from $Q_i$ to $Q_f$ in time $t$:
\be\label{PI}
\mathcal{P}(Q_f, t|Q_i) = e^{-\frac{G(Q_f)-G(Q_i)}{2 k_B T}}~ \int_{Q_i}^{Q_f}\mathcal{D}Q ~e^{-S_{eff}[Q]},\ee
where
\be
S_{eff}[Q]=\int_0^t d\tau \left(\frac{\dot Q^2}{4 D_Q} + V_{eff}[Q]\right)
\ee
 is called the effective action and 
 \be
 V_{eff}(Q) = \frac{D_Q}{4 (k_B T)^2}(|G'(Q)|^2- 2 k_B T G''(Q))
 \ee
  is called the effective potential. The DRP equations result from analyzing the path integral  (\ref{PI}) in saddle-point approximation. The saddle-point paths (called the dominant reaction 
pathways)  are the functional minima of the effective action. Hence, they obey the equation of motion 
\be\label{eom}
\ddot Q= 2 D_Q V'_{eff}(Q)~
\ee 
 and conserve the effective energy 
$
E_{eff}\equiv \frac{\dot Q^2}{4 D_Q}- V_{eff}[Q].
$

As discussed in detail Ref.s~\cite{DRP2,DRPfluct} the saddle-point paths which are relevant  in the description of thermal
 activation are those
which leave the reactant and reach the product with (nearly) vanishing velocity.
This observation implies 
\be
E_{eff} \sim ~-V_{eff}(Q_i)~\sim \frac{1}{2 \gamma} G''(Q_i),
\ee
 where we have used the fact the initial configuration $Q_i$ is in the vicinity of a free energy minimum.
On the other hand, outside the (meta-) stable thermodynamical states, the effective potential is dominated by its force contribution, 
\be
V_{eff}(Q )\sim 1/(4k_BT \gamma_Q) |G'(Q)|^2.\ee  
Hence,  in the transition region $E_{eff}/V_{eff}(Q)\sim o( k_B T)$.
 
The definition of effective energy immediately yields 
the time at which any given intermediate value $Q$ of the CC, located between the reactant $Q_R$ and the product $Q_P$, is visited by during a reaction~\cite{DRP0, DRP2}: 
\be\label{time}
t(Q) = \int_{Q_R}^{Q} \frac{dQ}{\sqrt{4D_Q(E_{eff}+V_{eff}[\bar Q])}}.
\ee
This equation can also be used compute the time at which the microscopic configurations in the dominant trajectories are reached, by imposing  $t(x)\simeq t(Q(x))$. In particular, Eq. (\ref{time}) provides an estimate of the TPT, which is obtained simply by setting $Q=Q_P$ and $E_{eff} \sim ~-V_{eff}(Q_R)$.

\subsection{Transition path time for crossing a harmonic barrier}

It is useful to discuss the TPT calculation within the simple harmonic approximation of a single free-energy barrier, which allows for an analytic treatment. 
In this case, the effective potential $V_{eff}(Q)$ is also a harmonic function and
reads
\be
V_{eff}(Q)\simeq  \frac{\alpha^2}{4 k_B T \gamma_Q} \cdot(Q-Q_{TS})^2 + \frac{\alpha}{2 \gamma_Q},
\ee
where $Q_{TS}$ identifies the transition state and $\alpha\equiv G''(Q_{TS})$.

In an harmonic barrier, the transitions which involve overcoming 
of an energy barrier $\Delta G$ are those initiated by a point $Q_i$ such that $|Q_{TS}-Q_i|= \sqrt{\frac{2 \Delta G}{\alpha}}$. 
Hence,  Eq. (\ref{time}) immediately gives 
\be
t_{TPT}
= \frac{\gamma}{\alpha} \ln\left[\frac{\left( \sqrt{2 \Delta G \alpha}+  2 \sqrt{k_B T[ E_{eff} 
+ \frac{\alpha}{2} (  1+ \frac{\Delta G}{ k_B  T})}]\right)^2}{2 k_B T(\alpha + 2 E_{eff})}\right]
\ee
Finally, retaining only the leading-order in the expansion in powers of the thermal energy $k_B T$ and recalling that $E_{eff}/V_{eff}\sim o(k_BT)$
we arrive to the final simple result,
\be
t_{TPT} = \frac{\gamma}{\alpha} \ln\left[4 \Delta G/(k_B T)\right],
\ee
which is close to the estimate obtained by Szabo, $t_{TPT}\simeq\frac{\gamma}{\alpha} \ln[3 \Delta G/(k_B T)]$. 
It should be emphasized however that Eq. (\ref{time}) generalizes this estimate beyond the harmonic approximation and the leading-order in the 
low-temperature expansion.

\section{Illustrative applications and Validation}
\label{tests}

For illustration and validation purposes, in the remaining of this work we apply and test our method  to characterize two reaction of increasing complexity. We begin by considering a simple two-dimensional toy model which can be straightforwardly implemented and for which analytic solution for the potential of mean-force exist. 

Next, we use our method to study a protein folding reaction within a coarse-grained representation of the polypeptide chain. In such a model, the folding reaction can be simulated directly by integrating the equation of motion and the results can be used to assess the accuracy of our method.
 
 \subsection{Diffusion on two-dimensional funneled potential}
\begin{figure}[t!]
\begin{center}
\includegraphics[width= 5cm]{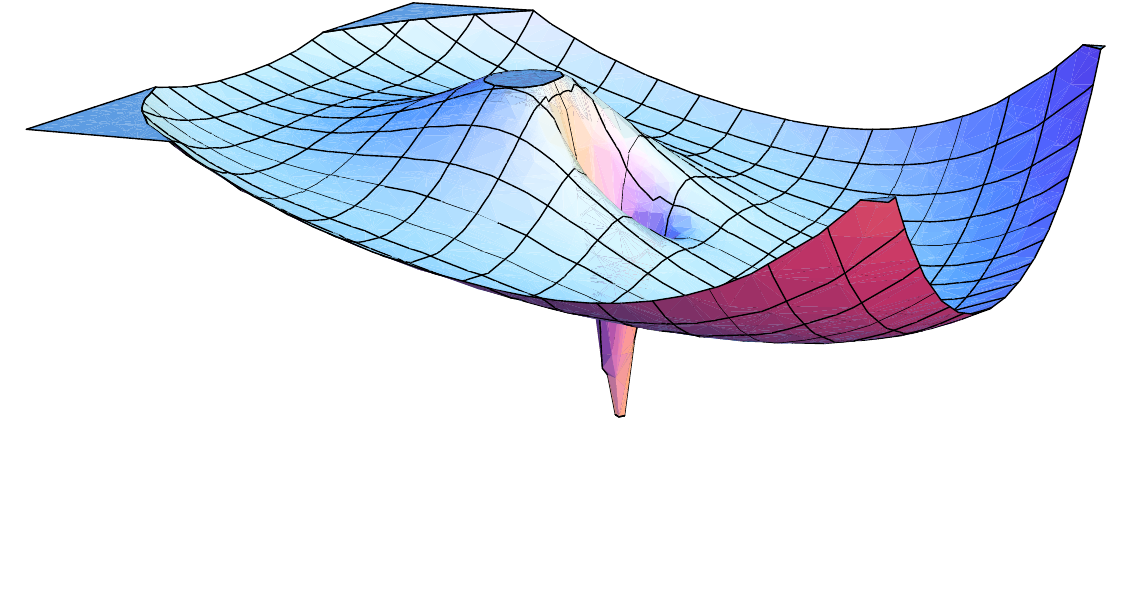}
\caption{(Color online) The energy surface of the two-dimensional toy model used to validate the method.}
\label{surface}
\end{center}
\end{figure}

We consider the over-damped Langevin diffusion of a point-particle in the two-dimensional funneled energy surface  shown in Fig.\ref{surface}, which is given by the potential:
\be
U(x,y) &=& \frac{- A_1 \sigma_1^2}{((x^2+y^2)+\sigma_1^2)^2} +  \frac{A_2 \sigma_1^2}{((x^2+y^2)+\sigma_2^2)} \nonumber\\
&& + \omega^2 (x^2+y^2)^2 + B \sin^2\left(\frac{\phi}{2}\right),\quad \phi = \arctan\left(y/x\right),\nonumber\\
\ee
with $A_1= 20$, $A_2=10$, $\sigma_1=1$, $\sigma_2=5$, $\omega=0.02$ and $B=10.$ As shown in Fig. \ref{surface} this model 
contains a stable  state at the origin and meta-stable state at some finite distance from the bottom of the funnel and $\phi\simeq 0$.  For $k_B T=1$, the barrier-crossing transition from the meta-stable to the stable state is thermally 
activated. The only slow coarse coordinate in this system
is the distance of the particle from the origin, $R=\sqrt{x^2+y^2}$ and the dominant reaction pathways are straight lines connecting the different initial conditions in the meta-stable 
state to the origin.   By contrast, a typical Langevin trajectory spends a large time in the metastable state, performs a barrier crossing transition and eventually lands in the stable state. 

The mean first-passage-time through the transition state, obtained from the Langevin simulations is $\langle t_{FPT} \rangle_{MD} =  532 \pm 40$
(units in which $\gamma\equiv1$).
The  mean TPT can be calculated by measuring the length of an ensemble of  Langevin trajectories which are started at the edge of the reactant
--arbitrarily defined by the condition $R_R= 5$-- and are terminated once the particle 
reaches the edge of the product-- identified by the condition $R_P=0.5$. Accumulating statistics 
only on the trajectories which do not visit the reactant before reaching the product we find $\langle t_{TPT}\rangle_{MD}= 2.65\pm 0.02$. 
 \begin{figure}[t!]
\begin{center}
\includegraphics[width=8cm]{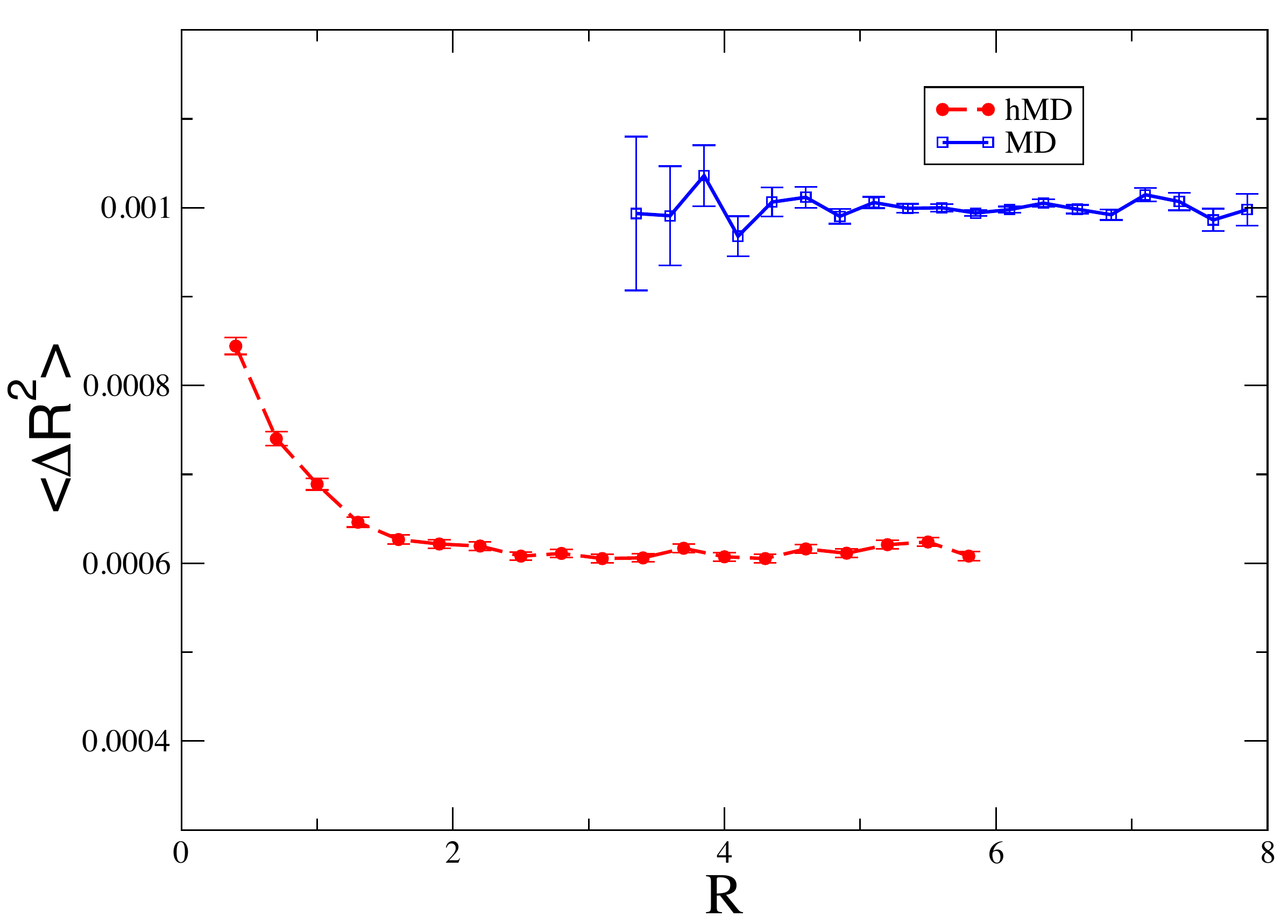}\\
\includegraphics[width=8cm]{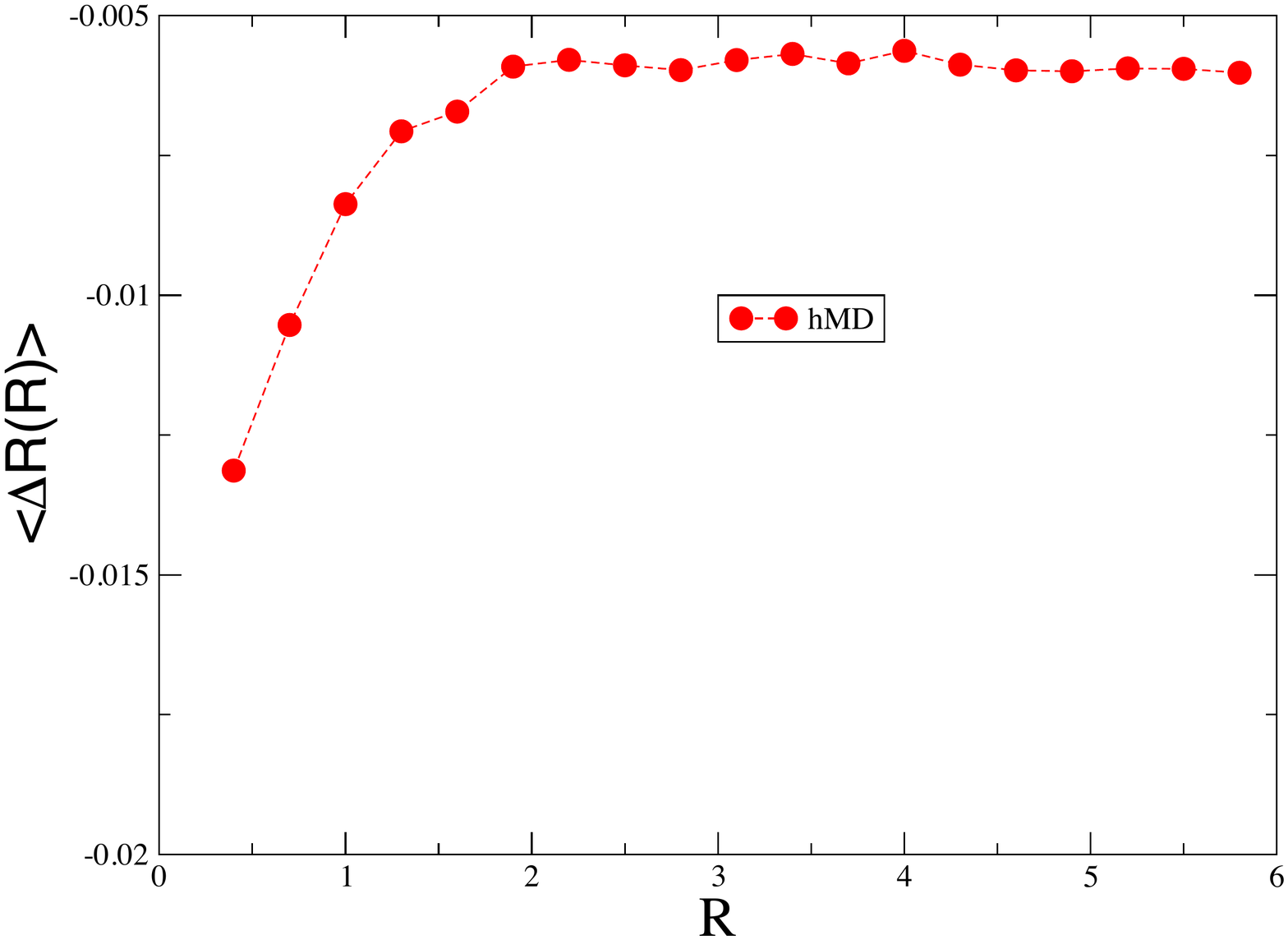}
\caption{(Color online) Upper panel: $\langle \Delta R^2(R )\rangle$ in the toy model, evaluated in hMD (circles) and MD (squares) simulations. Lower panel: 
  $\langle \Delta R(R )\rangle$ in hMD simulations. All quantities are in computer units.  }
\label{DRDR2}
\end{center}
\end{figure}

We now use Eq.s (\ref{main1}) and (\ref{main2}) to reconstruct the free energy landscape as a function  of the CC $z=R$. This is done by running 
hMD simulations which bias the dynamics towards smaller and smaller distances from the origin, according to the algorithm given in Eq. (\ref{hMD}). With a hindering coefficient $\xi=2$, 
generating a barrier crossing event requires simulating a time interval of about $0.4$, which is a factor $10^3$ times smaller than the mean-first-passage time. 
 hMD trajectories are functionally close to the dominant reaction pathway, i.e. to the straight radial line with $\phi=0$.       

According to our method, the first step towards reconstructing the free-energy surface
 consists in evaluating the friction coefficient for the CC used in the hMD, by means of Eq. (\ref{main2}). 
 Fig. \ref{DRDR2} shows   $\langle \Delta R^2(R )\rangle$ evaluated in the hMD simulation with
an elementary time interval  $\Delta t = 0.0005$. 
As predicted by  Eq. (\ref{main2}), this curve is flat almost everywhere. A weak dependence on $R$ is observed for $R\lesssim 2$, and 
is due to the fact that in the high-force region inside the funnel, gradient-dependent corrections to Eq. (\ref{main2}) which are higher order in $\Delta t $
 become relevant.  From a fit of the flat region, knowing that $\xi=2$, one obtains the correct result $\gamma_R\simeq 1$. 
To assess the validity of this calculation, in Fig. \ref{DRDR2} we also show the same average evaluated in standard (i.e. unbiased) MD simulations. 
According to Einstein's law  this average should be equal $\frac{2 \Delta t k_B T}{\gamma_R}$, which allows to confirm the result obtained from hMD.

\begin{figure}[t!]
\begin{center}
\includegraphics[width= 8cm]{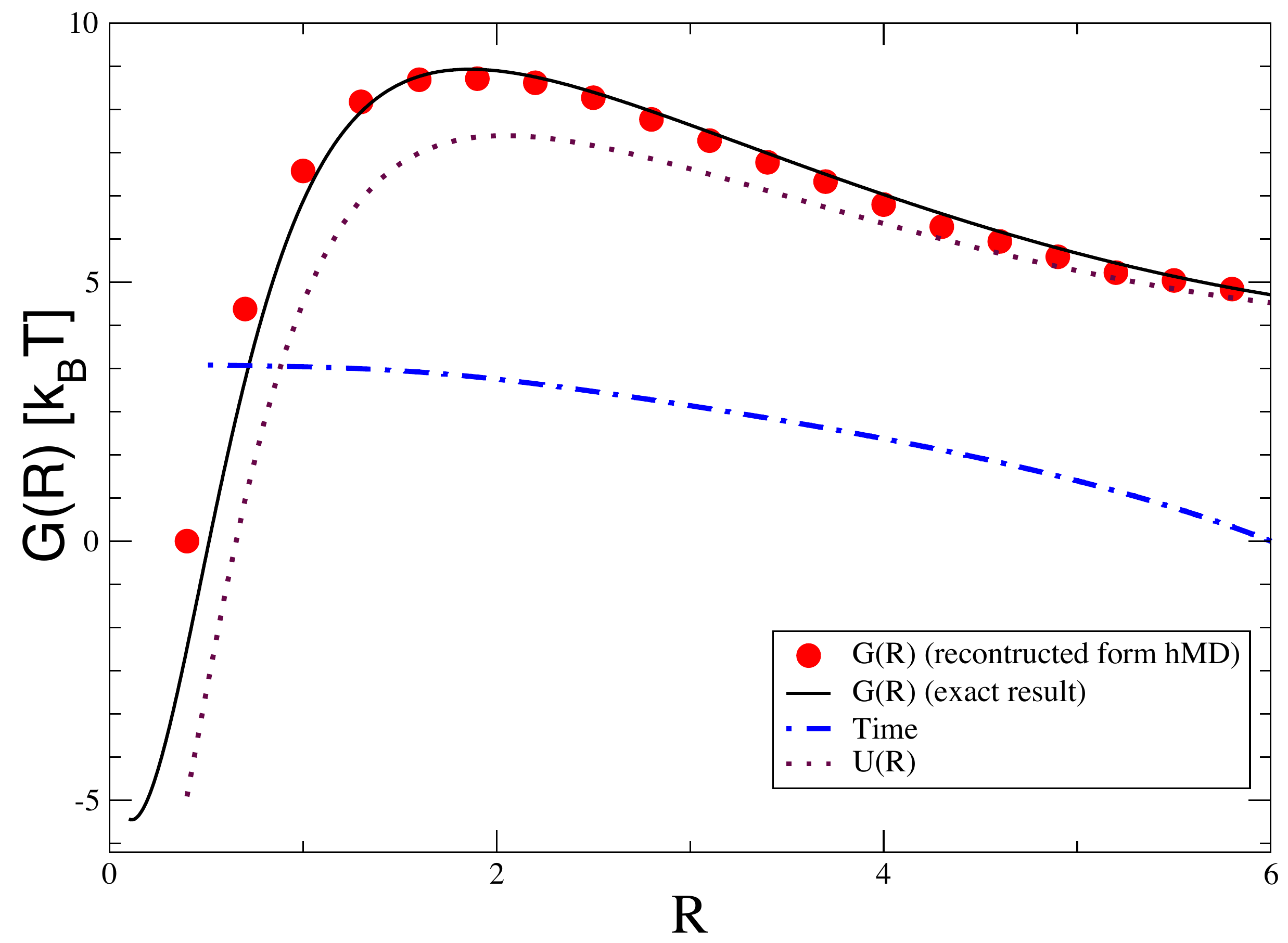}
\caption{ (Color online) Comparison between exact (solid line) and calculated (circles) free-energy profile as a function of the CC $R$. The dotted line represents the potential 
energy as a function of the coordinate $R$, evaluated along the radial line with polar coordinate $\phi=0$. The 
dot-dashed line shows the DRP time at which each value of $R$ is assumed.}
\label{GR}
\end{center}
\end{figure}

 Finally, knowing $\xi$ and having determined $\gamma_R$, it is possible to use Eq. (\ref{main1}) to extract the mean-force $G'(R )$ from the average displacement
  $\langle \Delta R( R)\rangle$ shown in Fig.~\ref{DRDR2},   hence to reconstruct $G( R)$. The calculated free-energy is shown in 
 Fig. \ref{GR}, where it is compared with the exact analytic result, $G(R ) = U(R ) -  k_B T \log \frac{R}{R_0}$,  -- here, $R_0$ is the  arbitrary reference point---.
The agreement between the two curves is quantitative.

The TPT estimated using the DRP equation (\ref{time}), setting the $E_{eff}=-V_{eff}(R_m)$ --where $R_m$ is the minimum free-energy distance
 in the meta-stable state---  gives $t_{TPT}^{DRP}\simeq 3.4$.
    This number is not far from the  average $\langle t_{TPT}\rangle_{MD}= 2.65\pm 0.02$, obtained by MD simulations. 
  In contrast, Szabo's formula, which relies on the harmonic approximation and low-temperature expansion,  gives $t_{TPT}^{Sz}=0.8$, which is off by a factor 3. 
This discrepancy suggest that temperature effects and specific curvature of the 
 energy surface at the transition state can give significant corrections to the TPT.

\subsection{The folding of a poly-petide chain} 

\begin{figure}[t!]
\begin{center}
\includegraphics[width= 8 cm]{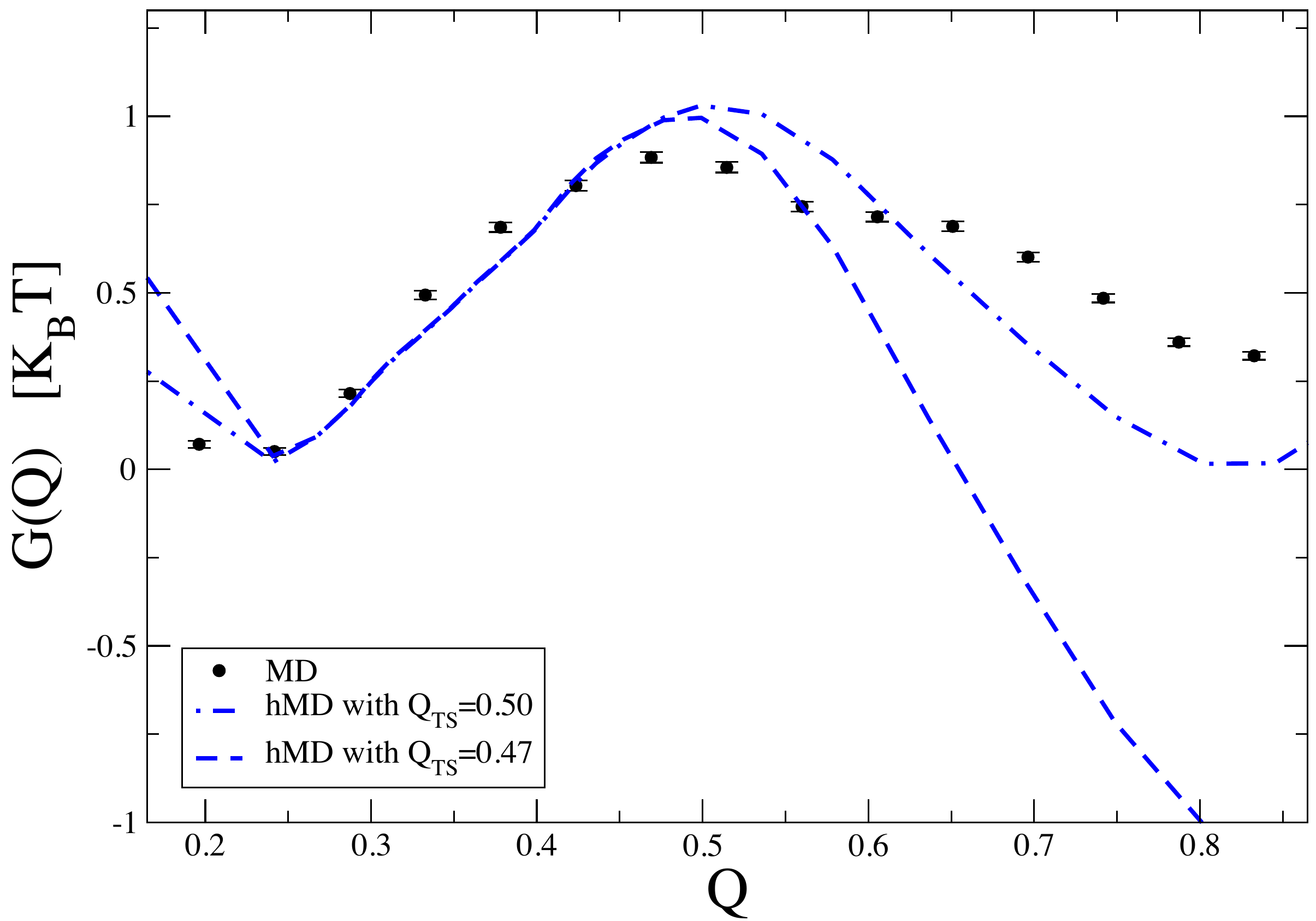}
\caption{ (Color online) Potential of mean-force as a function of the fraction of native contacts for the C-terminal of protein GB1 (the  native structure shown in the insert). }
\label{Results1}
\end{center}
\end{figure}

To further assess the accuracy and computational efficiency of our method,  we apply it to study a conformational reaction which resembles most of the difficulties which are encountered in  macromolecular systems: the folding of the 16 amino-acid C-terminus of protein GB1, whose native state is shown in the inset of Fig.\ref{Results1}.

To allow for a direct comparison with the result of standard MD simulations, we adopt the coarse-grained representation  introduced in Ref. \cite{DRPlonardi}. In such an approach,  the explicit degrees of freedom individual amino-acids and the energy function is a sum of pair-wise interactions: 
\be
U&=& \frac{1}{2} \sum_k  k (x_{k+1}-x_k|-a)^2+ \sum_{i<j} 4 \epsilon \left[A_{ij} \left(\frac{\sigma}{|x_{j}-x_i|}\right)^{12} \right.\nonumber\\
&&\left.- (G_{ij}+B_{ij})~\left(\frac{\sigma}{|x_{j}-x_i|}\right)^{6}\right]
\ee
where $x_i$ denotes the position of the $i-$th residue, $a=0.38$~nm, $k=3000 ~\textrm{kJ mo}l^{-1} \textrm{nm}^{-2}$, $\epsilon= 5 ~\textrm{kJ mol}^{-1}$, and $\sigma=0.3~$nm. 
$G_{ij}$ is the matrix of native contacts, i.e. $G_{ij}$  is set to $1$ if the distance between the residues $i$ and $j$ in the crystal native conformation is less than $0.65$~nm, and  $0$ otherwise (Go-type model, \cite{go}). The coefficients $A_{ij}$ and $B_{ij}$ are introduced in order  to account  for the hydro-philic (-phobic) character of the amino-acids, as in the  so-called HP  model \cite{HPmodel}. 

Namely, we set
\begin{itemize}
\item  $A_{ij}=1 $ and $B_{ij}=1 $, for pairs in which both amino-acids are  hydrophobic
\item $A_{ij}=\frac{2}{3}$ and $B_{ij}=-1$,  for  pairs in which one of the amino-acids is polar
\item $A_{ij}=1$ and $B_{ij}=0$ if one of the residues is GLY, which is hydrophobically neutral.
\end{itemize}
Such non-native interactions introduce ruggedness in the energy landscape, making simulations more challenging than in the purely native-centric model. 

 \begin{figure}[t!]
\begin{center}
\includegraphics[width= 7.5 cm]{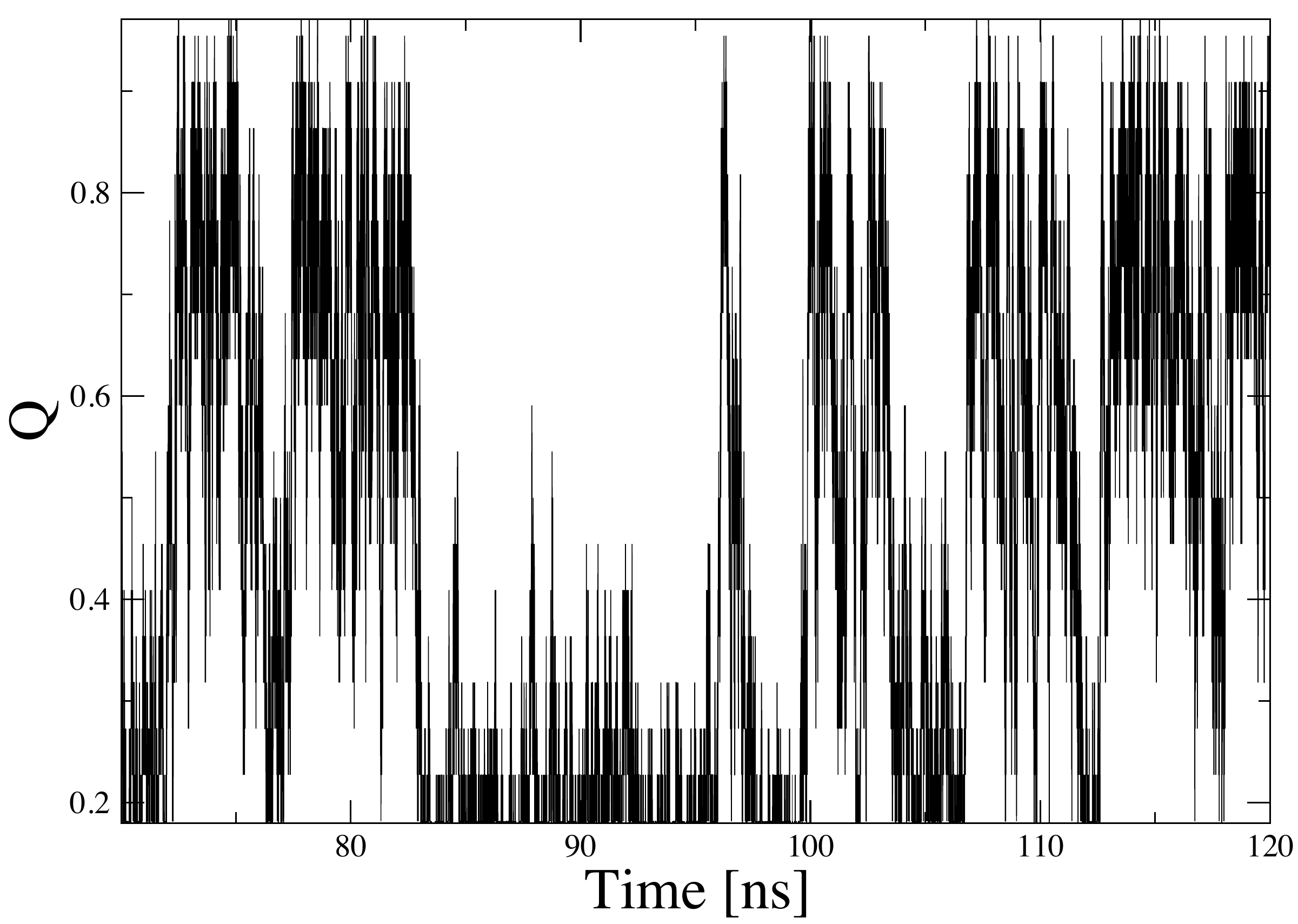}
\includegraphics[width= 7.5 cm]{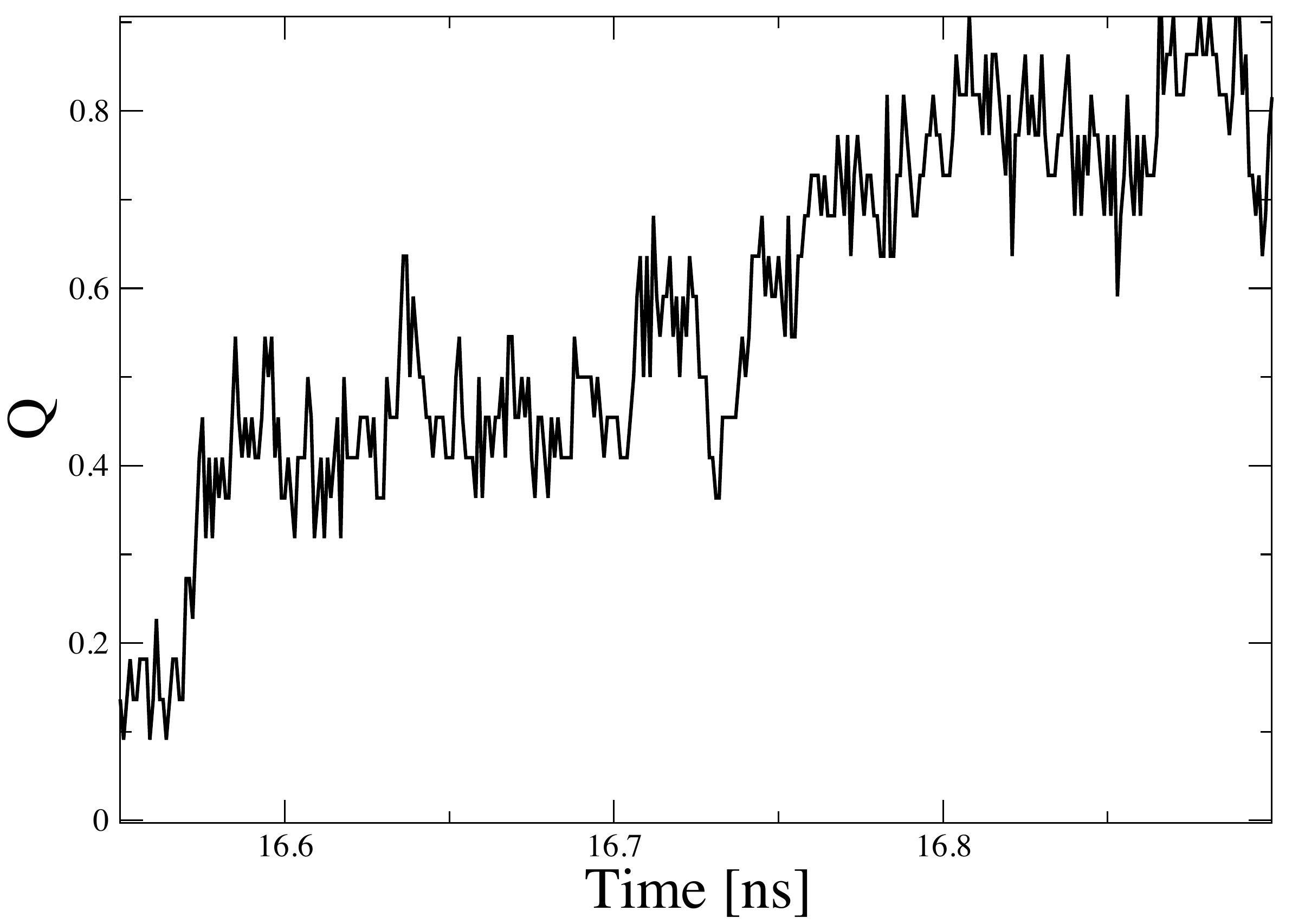}
\caption{ (Color online) The time evolution of the fraction of native contacts $Q$ of the poly-peptide obtained from  a long Langevin simulation with reversible folding-unfolding events (upper panel). 
The lower panel shows in detail the evolution of this variable along a typical folding event. The integration of the (Ito)  Langevin equation (\ref{langevin}) was performed at the nominal temperature of $T=200$~K,  with an integration time step $\Delta t=0.01$ps and a viscosity $\gamma=2000$~amu~ps$^{-1}$.    }
\label{MDx}
\end{center}
\end{figure}

The time evolution for the fraction of native contacts $Q$ (which is a commonly adopted reaction coordinate for protein folding) over a long Langevin trajectory is shown in Fig.~\ref{MDx}. 
The curve shown in the upper panel is compatible with a two-state system,  separated by a single low free-energy barrier. This fact is confirmed by the points shown in Fig.~\ref{Results1}, which represent the potential of mean-force for this system as a function of $Q$ obtained from a frequency histogram of the same trajectory.  The lower panel of Fig. \ref{MDx} shows the evolution of the fraction of native contacts along a  typical folding event.  The average transition path time for folding reactions  was found to be $\tau^{MD}_{TPT}=(0.50 \pm 0.05)$~ns.

Let us now discuss the calculation of the potential of mean-force using the hMD algorithm.  We performed 800 independent short hMD simulations with an hindering coefficient of $\xi=3$, starting from the fully stretched configuration. The trajectories were biased using a CC which counts the number of native contacts: 
\be
z= 1-\frac{1}{z_n}\sum_{i<j} (C(x_i, x_j) - G_{i j})^2,
\ee
where $z_n= \sum_{i<j} G_{i,j}$,
\be
C(x_i, x_j)=\frac{1-(|x_i-x_j|)/r_0)^6}{1-(|x_i-x_j|/r_0)^{10}},
\ee
and $r_0=0.7$~nm.  
This biasing  biasing CC was shown to be very efficient in guiding ratchet-and-pawl protein folding simulations~\cite{ratchetMD, DRPpnas}. 

\begin{figure}[t!]
\begin{center}
\includegraphics[width= 7.5 cm]{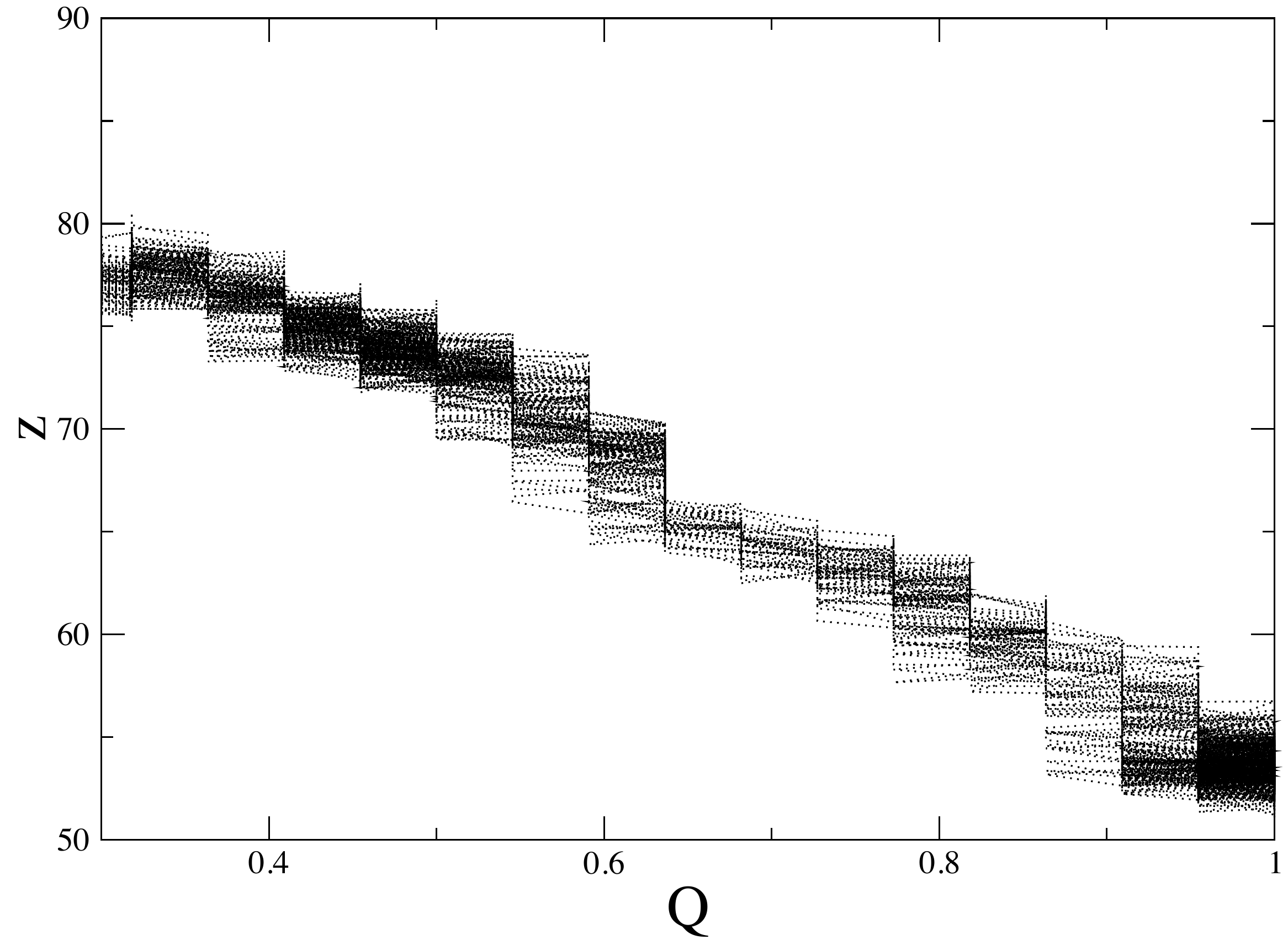}
\caption{ (Color online) Correlation between the ratchet variable $z$ and the fraction of native contacts $Q$ along the hMD trajectories used to reconstruct the free-energy surface.}
\label{zvsQ}
\end{center}
\end{figure}

In order to ease the physical interpretation of the results, we choose to reconstruct the free-energy surface as a function of the fraction of native contacts $Q$, rather than of the biasing variable $z$. We recall that, in order for the same hMD procedure to be transferable from the biasing coordinate to a reaction coordinate, the two variables should be directly proportional. Unfortunately, in general, this is not the case for $Q$ and $z$, as it is evident from the fact that the biasing variable can be increased not only by forming a native contact, but also by breaking a non-native contact. 
However figure \ref{zvsQ} shows that, while the linear correlation between $z$ is $Q$ can be violated in general, it is actually respected to good accuracy along the hMD folding trajectories, hence allowing the application of Eq.s (\ref{main1}) and (\ref{main2}). 

From the ensemble of hMD trajectories we have evaluated the average and average-square displacements of the fraction of native contacts, $\langle \Delta Q(Q) \rangle$ and $\langle \Delta Q^2(Q) \rangle$. 
The most straightforward way to reconstruct $G(Q)$ and compute $\gamma_Q$ from these averages consists in using Eq. (\ref{main2}) to 
fit $\gamma_Q$ (hence compute the diffusion coefficient $D_Q= k_BT/\gamma_Q$~), and then insert this value into Eq. (\ref{main1}) to extract  $G'(Q)$.  
However,  in the case of the present system,  we have observed  that higher-order corrections to Eq. (\ref{main2}) introduce
some modulation in $\langle \Delta Q^2(Q)\rangle$, which spoil the accuracy of an estimate of  $\gamma_Q$ based on a constant fit. We have therefore developed a different protocol to evaluate $G(Q)$ from the hMD averages,  inspired by   observation that the hindering of the dynamics should be suppressed once the system 
crosses the TS. Indeed, beyond this point, the molecule is rapidly and spontaneously relaxing to the product state, hence we expect that the mean 
displacement $\langle \Delta Q(Q) \rangle$  evaluated in hMD trajectories should undergo a rapid increase at the TS.   

\begin{figure}[t!]
\begin{center}
\includegraphics[width=7.5 cm]{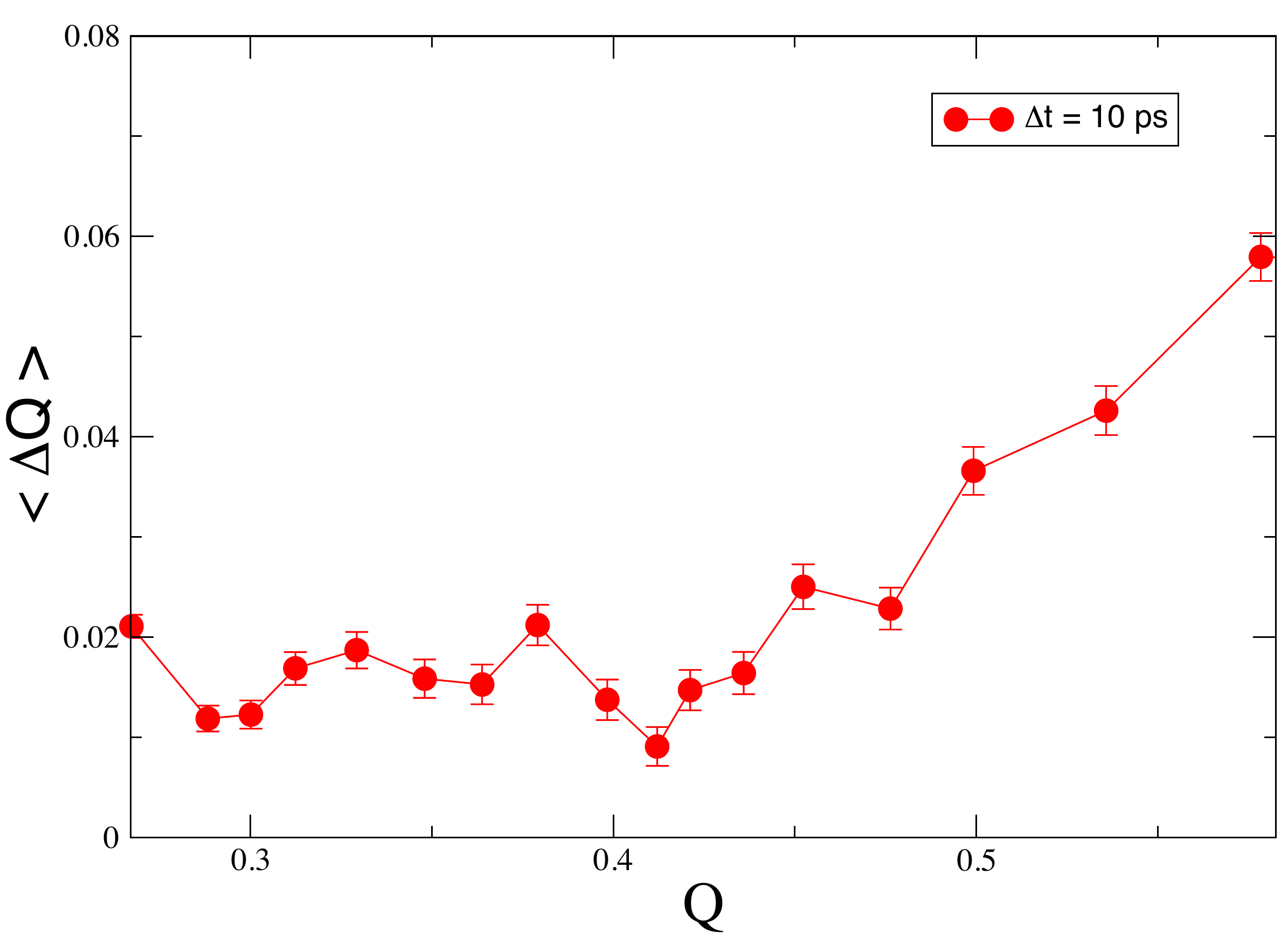}
\caption{ (Color online) Average displacement of the fraction of native contacts $\langle \Delta Q \rangle$ evaluated over a time interval $\Delta t= 10$~ps in the hMD simulation,  as a function of the fraction of native contact $Q$. }
\label{Results2}
\end{center}
\end{figure}

Fig. \ref{Results2}  shows the calculated $\langle \Delta Q(Q) \rangle$ ---evaluated over the  time interval $\Delta t=10$~ps---, which displays a steep increase in the region $0.47 \lesssim Q\lesssim 0.50$, which represents our estimate for the location of the TS. Using $G'(Q_{TS}) =0$, Eq. (\ref{main1}) leads to the estimate $\gamma_Q= (3200 \pm 1000)$~amu~ps$^{-1}$. Once this parameter has been fixed,  Eq. (\ref{main1}) yields $G'(Q)$ in all other points.  

The  results for the potential of mean-force are reported in Fig.~\ref{Results1} 
and show that the hMD algorithm is able to identify the two-state character of the reaction kinetics and gives a free-energy barrier to fold which is in very good agreement with  the results of MD simulations. On the other hand, the prediction for the free-energy profile is much less accurate in the region from the TS to the native state. 
This fact is  expected, since in the hMD approach the free-energy is obtained  by comparing local fluctuations of the velocity in the presence and absence of the hindering. Clearly, beyond the TS the hindering is  suppressed and the method becomes inaccurate. This does not represent a problem, since the unfolding free-energy barrier may in principle be calculated by the same algorithm using a different bias which drives the chain from the native to the denatured state. 
  \begin{figure}[t!]
\begin{center}
\includegraphics[width=7.5 cm]{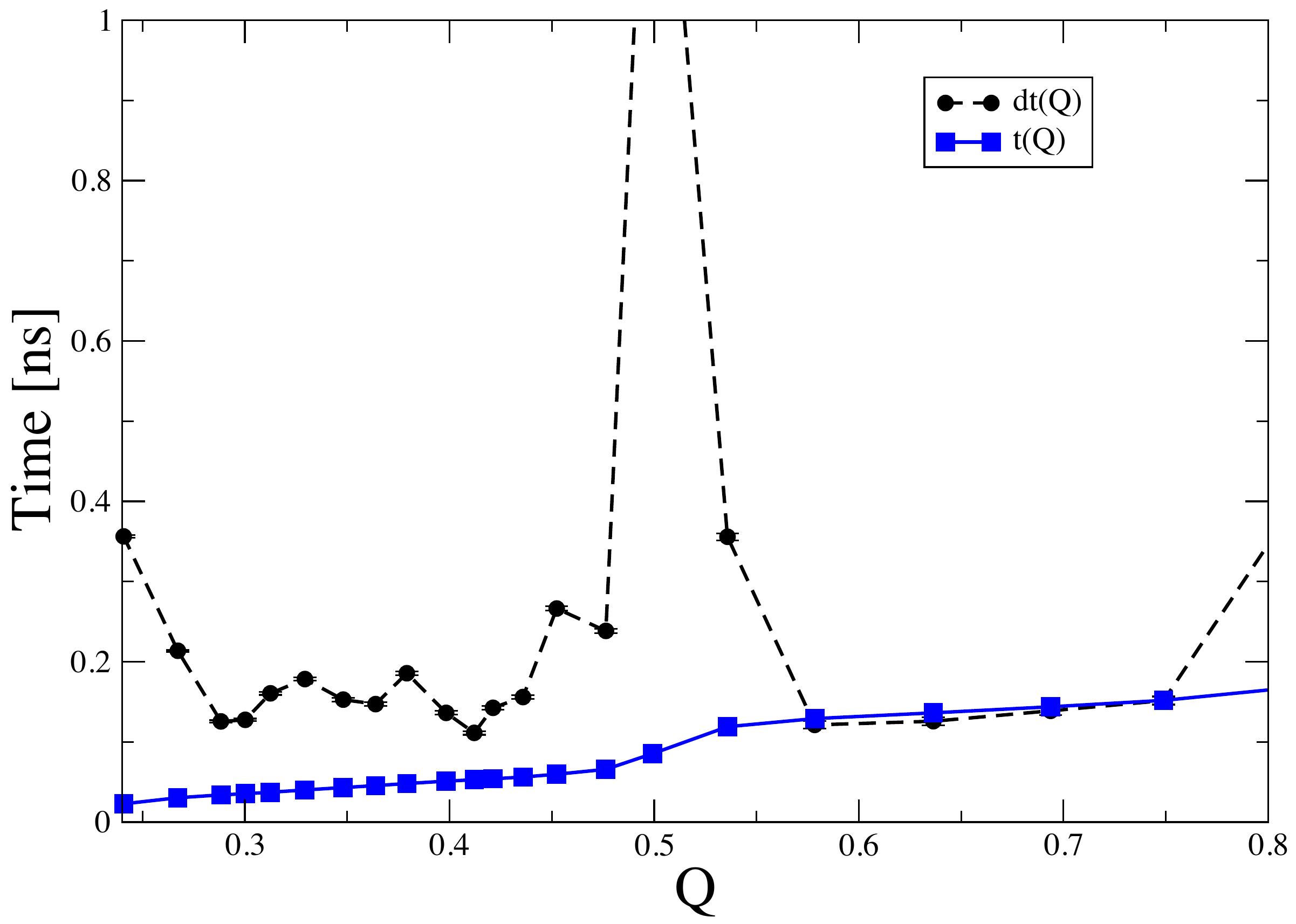}
\caption{(Color online)  The time at which each value of the CC is visited, computed from the DRP equation (\ref{time}) using the calculated $G(Q)$ and $\gamma_Q$. The dotted line shows the residence time $dt(Q)$, i.e. the integrand of Eq. (\ref{time}). The initial condition is represented by the point on the left of the plot. }
\label{Results3}
\end{center}
\end{figure}

 The time at which each value of the CC is visited, computed from the DRP equation (\ref{time}) using the calculated $G(Q)$ and $\gamma_Q$  is shown in Fig. (\ref{Results3}). . 
 The dotted line shows the residence time $dt(Q)$, i.e. the integrand of Eq. (\ref{time}). We note that the residence time is longer 
 in the region around $Q\simeq 0.5$. The same feature is observed in the unbiased Langevin simulations (cfr. the lower panel in Fig.\ref{MDx}). 
 The DRP estimate for the total TPT is $\tau_{TPT}\simeq 0.45\pm 0.15$~ns, again in good  agreement  with the results of the MD simulations, $\tau^{MD}_{TPT}=(0.50 \pm 0.05)$~ns.  
We emphasize  that the number of hMD trajectories needed to reconstruct $G(Q)$ and compute $\tau_{TPT}$ is of the same order of those which have been generated in our previous atomistic DRP simulations of protein folding. 

 \section{Conclusions}\label{conclusions}
 
In this work we have introduced an accelerated MD which allows to  compute   the free-energy profile $G(Q)$ and the diffusion coefficient $D_Q$ which describe the stochastic dynamics of a previously determined \emph{slow} collective variable~$Q$. By applying the DRP formalism we have shown that, once $G(Q)$ has been
calculated, it is straightforward to obtain an estimate of the TPT, which holds up to logarithmic corrections.

The main advantages of the present method are (i) that the acceleration of the dynamics is not generated by any external force and (ii) that the systematic errors 
introduced in order to accelerate the overcoming of the free-energy barriers can be analytically computed, hence corrected for. 

A few  remarks on the limitations of this method are in order.  First, we emphasize that it relies on the possibility of identifying a single slow reaction coordinate for the macromolecular system (a list of methods developed to this purpose can be found e.g. in Ref.s \cite{RC1, RC2, RC3, RC4, RC5, RC6}).
If a poorly chosen reaction coordinate is used, the free-energy profile is not expected to capture the correct rate-limiting barrier, and reaction rate calculation will
 be exponentially inaccurate.  

 In general, it is not always possible to identify a single slow coarse variable (see e.g. in maze problem~\cite{maze}). In these cases, our method is expected to miss substantial entropic contributions to the free-energy. Finally, we have found that the free-energy reconstruction is much less accurate in the region connecting the transition state to the product. 
 
All such limitations significant impact on the accuracy of free-energy profile calculations  and reaction kinetics. On the other hand, the calculation of the TPT and of the time intervals between consecutive frames in  DRP reaction paths are expected to be much more reliable, as the depend only logarithmically on the free-energy.
 

\section*{Acknowledgment}{PF acknowledges stimulating discussions with G. Tiana and A. Szabo. }
\appendix

\end{document}